\documentclass[amsmath, amssymb, aps,twocolumn,noshowpacs,superscriptaddress, pra, nofootinbib]{revtex4-2}

\usepackage[hidelinks]{hyperref}% add hypertext capabilities
\usepackage[utf8]{inputenc}%Umlaute in bibtex

\usepackage{graphicx}% Include figure files
\usepackage{dcolumn}% Align table columns on decimal point
\usepackage{bm}% bold math
\usepackage[capitalise]{cleveref}%nice references with Cref
\usepackage{bbold} %for identity matrix
\usepackage{color}
\usepackage{braket}
\usepackage{dsfont}

\newcommand{\one}{\mathbb{1}}

\newcommand{\tr}{{\rm Tr}}
\newcommand{\ketbra}[2]{\ket{#1}\bra{#2}}
\newcommand{\1}{\mathds{1}}

\newcommand{\affilITP}{Institute for Theoretical Physics, ETH Z\"{u}rich, CH-8093 Z\"urich, Switzerland.}
\newcommand{\affilKON}{Department of Physics, University of Konstanz, 78464 Konstanz, Germany.}

\begin{document}

\preprint{APS/123-QED}

\title{Scrambling-induced entanglement suppression in noisy quantum circuits}% Force line breaks with \\

\author{Lea Haas}\thanks{These two authors contributed equally to this work}\affiliation{\affilITP}\affiliation{\affilKON}
\author{Christian Carisch}\thanks{These two authors contributed equally to this work}\affiliation{\affilITP}
\author{Oded Zilberberg}\affiliation{\affilKON}

\date{\today}% It is always \today, today,
             %  but any date may be explicitly specified

\begin{abstract}
Quantum information scrambling is a process happening during thermalization in quantum systems and describes the delocalization of quantum information.
It is closely tied to entanglement, a key resource for quantum technologies and an order parameter for quantum many-body phenomena.
We investigate the effect of dephasing noise on a multi-qubit teleportation protocol that experimentally validated quantum information scrambling.
We find that while scrambling enhances information distribution, it is highly noise-sensitive, leading to decreased teleportation fidelity and an increase in the classical mixing of the quantum state.
Using negativity as a mixed-state entanglement measure, we identify two fundamentally different entanglement-scaling regimes: efficient entanglement generation under weak dephasing noise, and entanglement suppression under strong dephasing noise.
We show that in the latter, the teleportation consumes more entanglement than the scrambling is able to create.
Comparison with a SWAP-gate-based teleportation protocol confirms 
that the entanglement suppression is a consequence of the scrambling mechanism.
Our findings suggest that the information dynamics during thermalization is critically affected by dephasing noise, and confirm that in present-day noisy quantum devices, local information exchange is preferable over long-range information scrambling.
\end{abstract}

%\keywords{Suggested keywords}%Use showkeys class option if keyword
                              %display desired
\maketitle

%\tableofcontents

\section{Introduction}
How does a physical system relax to its equilibrium state? This process of thermalization is among the most fundamental in physics. Classical systems thermalize (irreversibly) when they are ergodic, i.e., when their macroscopic equilibrium state reflects all possible microscopic realization according to their Boltzmann probabilities~\cite{gallavotti1999statistical}. The question about how thermalization extends to the underlying quantum mechanical processes is highly complex as closed quantum systems evolve according to the linear (and therefore reversible) Schrödinger equation~\cite{mori2018thermalization}. Indeed, several quantum mechanical systems withstand thermalization, leading to effects such as many-body localization (MBL) in presence of disorder~\cite{PhysRev.109.1492,PhysRevB.75.155111,PhysRevB.82.174411,RevModPhys.91.021001,PhysRevE.102.062144}, or quantum (many-body) scars resulting from long-time memory of the initial state information~\cite{PhysRevLett.53.1515,serbyn2021quantum,PhysRevLett.132.021601}. For other quantum systems, however, even single energy eigenstates seem to be thermal, as conjectured by the eigenstate-thermalization hypothesis (ETH)~\cite{PhysRevA.43.2046,PhysRevE.50.888,rigol2008thermalization,PhysRevLett.105.250401}. The evolution of (non-)thermalizing systems is deeply connected to how local information gets distributed across the whole system, a process called information scrambling~\cite{Hayden_2007, sekino2008fast, shenker_stanford_2014,hosur_et_al_2016}.

Quantum information scrambling was initially considered in the context of black holes, which are the fastest scramblers in the universe~\cite{Hayden_2007,sekino2008fast}. When a black hole scrambles the information about a quantum state, the emitted Hawking radiation~\cite{hawking_1975} allows to decode this information and teleport the quantum state outside the event horizon. This thought experiment sparked the idea that successful teleportation can serve as a marker of quantum information scrambling~\cite{yoshida2017efficient,yoshida2019disentangling}. To explore this conjecture, a recent experiment mimicked a traversable wormhole~\cite{gao_et_al_2017, maldacena_et_al_2017} on a trapped ion quantum computer and verified quantum information scrambling using teleportation~\cite{landsman_verified_2019}.
To this end, the minimal teleportation protocol using one Bell pair (two maximally entangled qubits) and a Bell measurement (a joint measurement of two qubits in the Bell pair basis)~\cite{bennett_et_al_1996, bouwmeester_experimental_1997, nielsen_chuang_2010} was extended to include a unitary matrix simulating quantum information scrambling.
A crucial advantage of scrambling verification using teleportation is the stability of the teleportation protocol against environmental noise (decoherence). Other scrambling signallers such as out-of-time ordered correlators (OTOCs)~\cite{shenker2014black,maldacena2016bound,swingle2018unscrambling} generally decay during both scrambling and decoherence, and as such fail to distinguish between these two fundamentally different processes.
As the Hawking radiation decoheres black holes~\cite{demers_kiefer_1996,kiefer_2001,hsu_reeb_2009,arrasmith_et_al_2019,del_campo_takayanagi_2020}, it is essential to study thermalization using a quantity that is able to differentiate between scrambling and decoherence.

Another quantity of interest in the context of thermalization and information scrambling is the entanglement in the system.
While the quantum information scrambling is characteristic of the global thermalization process, the entanglement captures the local spread of quantum information~\cite{lieb_robinson_1972}, which can happen over fundamentally different time scales~\cite{Bohrdt_2017}.
Nevertheless, entanglement is key in the aforementioned teleportation protocol; the scrambling delocalizes the information across the system, whereas entanglement is the resource enabling the teleportation in the first place~\cite{bennett_teleporting_1993,bouwmeester_experimental_1997}.
Entanglement is also a key resource for quantum information processing~\cite{nielsen_chuang_2010, boixo_et_al_2018, neill_et_al_2018, arute_et_al_2019}, quantum metrology~\cite{PhysRevLett.96.010401,giovannetti2011advances}, quantum error correction~\cite{bennett_et_al_1996, cory_et_al_1998, schindler_et_al_2011, andersen_et_al_2020, krinner_et_al_2021}, and secure quantum communication~\cite{long_et_al_2007, hu_et_al_2016, zhang_et_al_2017}. Moreover, it serves as an order parameter for measurement-induced phase transitions of quantum systems coupled to detectors~\cite{li_et_al_2018,potter_vasseur_2022,fisher_et_al_2023,noel_et_al_2022,koh_et_al_2023,hoke_et_al_2023,carisch_romito_zilberberg_2023,carisch2024doesentanglementcarereadout}. That said, the experimentally realized teleportation protocol in Ref.~\cite{landsman_verified_2019} offers a versatile platform to study the interplay of fundamental effects such as scrambling, decoherence, and quantum measurement.

In this work, we systematically analyze the scrambling-based quantum teleportation protocol~\cite{landsman_verified_2019} in the presence of tunable dephasing noise. To this end, we consider the teleportation fidelity, the purity, and the mixed-state entanglement of the circuit. We find that teleportation can only be successful at highly efficient scrambling and weak dephasing. As expected, the dephasing turns the circuit increasingly classical, resulting in a loss of purity and fidelity. Intriguingly, the scrambling enhances this process and makes the circuit more vulnerable to dephasing, i.e., it distributes also classical correlations faster across the system.
For the entanglement, we find two different regimes: (i) for weak dephasing, the scrambling efficiently generates entanglement in the circuit, whereas (ii) for strong noise, the scrambling generates less entanglement than the Bell measurement at the end of the teleportation protocol consumes, i.e., we observe scrambling-induced entanglement suppression. A comparison to a teleportation circuit based on SWAP gates confirms that the regime of entanglement suppression is a consequence of the scrambling mechanism. Our findings show that quantum information scrambling can only delocalize information efficiently at weak noise; for noisy circuits, local exchange of information is preferable.

This work is organized as follows: In Sec.~\ref{sect:setup}, we describe the setup of the scrambling and SWAP-gate-based teleportation protocols and introduce the relevant quantities we use to analyze them.
In Sec.~\ref{sec: scrambling}, we discuss the scrambling circuit in more detail, while in Sec.~\ref{sec: comparison}, we compare our findings from the scrambling circuit to the SWAP-gate-based teleportation protocol.
Then, in Sec.~\ref{sec: total entanglement}, we consider the total entanglement generation and consumption in both circuits.
Finally, in Sec.~\ref{sec: conclusion}, we conclude our results and put them in perspective.
In the Appendix, we provide further details on the teleportation circuits and our numerical procedure.
\section{Setup}\label{sect:setup}

\begin{figure}[t]
    \centering
    \includegraphics{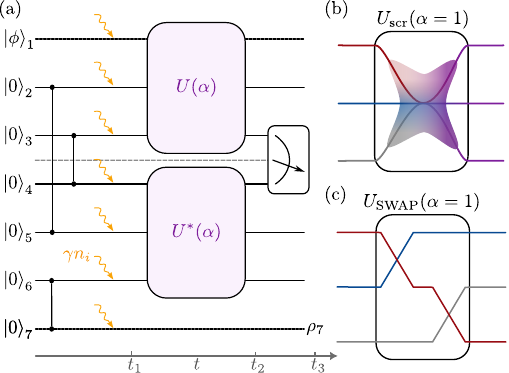}
    \caption{(a) A seven-qubit protocol teleporting the source qubit state $\ket{\phi}_1$ to the target qubit state $\rho_7$.
    The circuit is initialized with 3 Bell pairs (vertical lines) and each qubit is coupled to a dephasing environment with strength $\gamma$ [cf. Eq.~\eqref{eq:lindblad}].
    The teleportation is enabled by encoding with the 3 qubit unitary $U(\alpha)$ and decoding with its conjugate $U^*(\alpha)$, where $0\leq \alpha\leq 1$ is the transmission strength.
    This happens between times $t_1$ and $t_2$.
    A Bell measurement at time $t_3$ finalizes the circuit with heralded teleportation depending on the measurement outcome.
    (b) Scrambling realization~\cite{landsman_verified_2019} of the encoding unitary $U(\alpha)$ in (a) for maximal scrambling $\alpha=1$.
    Qubit-local information, indicated by the different colors of the incoming qubits, gets maximally scrambled across all three qubits, indicated by the single color purple.
    For $\alpha<1$, the distribution would not be even across the qubits.
    (c) SWAP realization of the encoding unitary $U(\alpha)$ in (a). For maximal swapping strength $\alpha=1$, the unitary consists of two consecutive SWAP gates. For $\alpha<1$, the swapping would be incomplete.}
    \label{fig:setup}
\end{figure}

We consider two realizations of quantum teleportation in an open 7-qubit quantum circuit [cf. Fig.~\ref{fig:setup}]: (i) a scrambling-based and a (ii) SWAP-gate-based teleportation protocol.
To teleport the state $\ket{\phi}$ from the source qubit 1 to the target qubit 7, both circuit realizations are initialized with 3 Bell pairs, whose entanglement acts as a resource to transport information.
Then, the information about the state of the source qubit is distributed across the circuit using a tunable 3-qubit unitary gate $U(\alpha)$, where $\alpha$ is a parameter controlling the distribution strength;
for $\alpha=0$, the information is not distributed ($U(0) = \one$), whereas for $\alpha=1$, the information is distributed perfectly (in the noiseless limit).
For the (i) scrambling circuit, $U(\alpha)=U_{\rm scr}(\alpha)$ is a 3-qubit scrambling matrix [cf. Fig.~\ref{fig:setup}(b)], and for the (ii) SWAP-gate based teleportation circuit, $U(\alpha)=U_{\rm SWAP}(\alpha)$ consists of two parametrized SWAP gates [cf. Fig.~\ref{fig:setup}(c)].
When we discuss the scrambling (SWAP-gate-based) circuit, we refer to $\alpha$ as the scrambling (swapping) strength.
The action of $U(\alpha)$ can be understood as an encoding of the local information of the state of qubit 1 across the circuit.
Consequently, simultaneously to the action of $U(\alpha)$, the distributed information is decoded using the conjugate $U^*(\alpha)$.
The final step is a heralded Bell measurement of qubits 3 and 4, where we postselect those runs where the qubits are found to be in the Bell pair $\ket{\Phi}^+=\frac{1}{\sqrt{2}}(\ket{00} + \ket{11})$.
In a successful teleportation, the measurement projects the target qubit onto the desired state of the source qubit, $\rho_7=\ketbra{\phi}{\phi}$.

The scrambling circuit (i) mimics a traversable wormhole~\cite{gao_et_al_2017, maldacena_et_al_2017} and the teleportation realizes the decoding of the emitted
Hawking radiation from the Black hole in the aforementioned thought experiment~\cite{Hayden_2007, yoshida2019disentangling, yoshida2017efficient}.
This circuit was experimentally realized in Ref.~\cite{landsman_verified_2019}.
Notably, the scrambling by the unitary $U_{\rm scr}(\alpha)$ delocalizes the initial state information across the whole circuit, enabling teleportation by performing a Bell measurement on any of the qubit pairs (1,6), (2,5), or (3,4)~\cite{yoshida2019disentangling, landsman_verified_2019}.
An explicit form of the scrambling gate is given in Appendix~\ref{sect:appendix_circuits}.
On the other hand, the SWAP-gate teleportation realization (ii) with $U_{\rm SWAP}(\alpha)$ locally exchanges the initial state information by consecutive applications of parametrized SWAP gates between neighboring qubits [cf. Fig.~\ref{fig:setup}(c) and Appendix~\ref{sect:appendix_circuits}].
Here, $\alpha$ continuously tunes the gates from no ($\alpha=0$) to perfect qubit exchange ($\alpha=1$).
Note that for $\alpha = 1$, this circuit simplifies to the minimal 3-qubit teleportation protocol using a single Bell pair~\cite{bennett_teleporting_1993, bouwmeester_experimental_1997,boschi_experimental_1998}.

In order to describe a realistic experimental setup, we consider continuous-in-time implementations of the Bell pair initialization, gate applications, and the Bell measurement, while coupling the qubits to local dephasing environments, see Fig.~\ref{fig:setup}(a) and Appendix~\ref{sect:appendix_numerical_details}.
As such, the dynamics of the system is governed by the Lindblad master equation~\cite{lindblad_1976},
\begin{eqnarray}
    \frac{d}{dt} \rho(t) = && - \frac{i}{\hbar} [H(\alpha,t), \rho(t)] \nonumber\\
    &&+ \gamma \sum_{i = 1}^{7} \left(n_i \rho(t) n_i - \frac{1}{2} \{\rho(t), n_i\} \right)\, , \label{eq:lindblad}
\end{eqnarray}
where $H(\alpha,t)$ is the time-dependent Hamiltonian describing the qubit manipulations and the density jump operators $n_i = \frac{1}{2} (\one + Z_{i})$, with $Z_{i}$ the Pauli-Z matrix on site $i$, encode the capacitive coupling  of the qubits to their dephasing environments.
To numerically evolve Eq.~\eqref{eq:lindblad} from time $t=0$ to $t=t_3$ after the Bell measurement, we discretize the time into bins, $t_3 = N\Delta t$ ($N\gg 1$), and for each time bin separately perform the unitary and the dissipative update, see Appendix~\ref{sect:appendix_numerical_details} for more details.
%The dephasing strength $\gamma$ is the inverse of the dephasing time $T_2 =1/\gamma$.

To understand the mechanisms behind how information is distributed in the two different teleportation protocols, we consider three key qualities of the circuit: (i) the success of the teleportation, (ii) the degree of quantumness of the final state, and (iii) the excess entanglement available after the Bell measurement.
We (i) quantify the teleportation success using the average teleportation fidelity, 
\begin{equation}
    \overline{F} = \overline{\bra{\phi} \rho_7 \ket{\phi}}\,, \label{eq:fidelity}
\end{equation}
where the overline denotes averaging over different initial states $\ket{\phi}$ [cf. Appendix~\ref{sect:appendix_numerical_details}] and $\rho_7=\tr_{1-6}\rho$ is the reduced state of the target qubit 7 after the protocol.
It holds that $0\leq F\leq 1$, where $F=0$ implies that the target state is orthogonal to the input state, and $F=1$ marks perfect teleportation, i.e., $\rho_7=\ketbra{\phi}{\phi}$.
If the target qubit is in a completely random state, $\rho_7=\1/2$, then $F=1/2$.
In order to (ii) quantify the quantumness of the final state, we consider its purity,
\begin{equation}
    \overline{\mathcal{P}} = \tr{\overline{\rho^2}}\label{eq:purity}\,.
\end{equation}
The purity is limited by $1 \leq \mathcal{P} \leq 1/d$, where $d = 2^7$ is the dimension of the system's Hilbert space.
The purity equals one only for pure (quantum) states $\rho=\ketbra{\psi}{\psi}$, and is $\mathcal{P}=1/d$ for a completely mixed (classical) state $\rho = \one/d$.
In our setup, the purity captures the amount of classicality induced by the interactions of the qubits with their environments.

Finally, the (iii) excess entanglement is the amount of available entanglement after the teleportation.
In order to quantify the entanglement, we employ the logarithmic negativity~\cite{vidal_computable_2002, plenio_logarithmic_2005}
\begin{equation}
\label{eq: log negativity}
    \overline{E}_\mathcal{N} = \overline{\ln (1 + 2 \mathcal{N})}\,, \quad \mathcal{N} = \frac{\sum_{i} |\lambda_i|-\lambda_i}{2}\,,
\end{equation}
where $\lambda_i$ are the eigenvalues of $\rho^{T_B}$, which is the partial transpose of the density matrix with respect to subsystem $B$, i.e.,
\begin{equation}
    \bra{i_A, j_B} \rho^{T_B} \ket{k_A, l_B} = \bra{i_A, l_B} \rho \ket{k_A, j_B}\,,
\end{equation}
and $\ket{i_A, j_B}$ are elements of a bipartite product basis.
Notably, the logarithmic negativity between two qubits in a (maximally entangled) Bell pair is $E_\mathcal{N}=1$, such that for pure states, it counts the entanglement equivalent in number of Bell pairs.
Furthermore, the logarithmic negativity~\eqref{eq: log negativity} is a mixed state entanglement measure and can therefore quantify the amount of entanglement also in the presence of dephasing noise.
In the following, we will consider both the entanglement for a bipartition between the encoding and decoding parties (i.e., for a cut between qubits 3 and 4, see Fig.~\ref{fig:setup}(a)) and the total entanglement given as the sum over the logarithmic negativities between each pair of neighboring qubits.

\section{Scrambling Circuit}
\label{sec: scrambling}
\begin{figure}[t]
    \centering
    \includegraphics{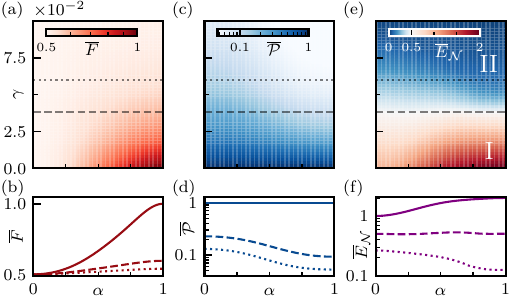}
    \caption{Fidelity $\overline{F}$, purity $\overline{\mathcal{P}}$, and negativity $\overline{E}_\mathcal{N}$ of the final state after the scrambling teleportation circuit [cf. Figs.\ref{fig:setup}(a) and (b) and \cref{eq:fidelity,eq:purity,eq: log negativity}, respectively] as a function of scrambling strength $\alpha$ and dephasing strength $\gamma$.
    (b,d,f) show the same data as (a,c,e) for specific dephasing strength cuts $\gamma \in \{0, 0.038, 0.06\}$ (solid, dashed, dotted).
    (a) Successful teleportation ($\overline{F}\lessapprox 1$) requires high scrambling and weak dephasing.
    Strong dephasing and/or weak scrambling leads to an unkown final state with fidelity $\overline{F}~\approx~1/2$.
    (b) At all dephasing strengths, the fidelity monotonously increases with the scrambling parameter $\alpha$. Fully successful teleportation ($\overline{F}=1$) is only possible for $\alpha = 1$ and in the absence of dephasing ($\gamma = 0$).
    (c) Increasing dephasing $\gamma$ leads to a more classical final state with low purity $\overline{\mathcal{P}}$.
    (d) The scrambling accelerates the decrease in purity at finite dephasing strengths $\gamma>0$. Without noise, the state is pure ($\overline{\mathcal{P}} = 1$) for all scrambling parameters $\alpha$.
    (e) In contrast to the fidelity and the purity, the logarithmic negativity between qubits 3 and 4 exhibits two qualitatively different regimes: (I) at weak dephasing, the entanglement increases with scrambling, while (II) at strong dephasing, the entanglement decreases with scrambling.
    (f) The two regimes (I) and (II) are separated by a critical error strength $\gamma = 0.038$, for which $\overline{E}_\mathcal{N}$ is approximately constant in $\alpha$.}
    \label{fig:scrambling}
\end{figure}

First, we consider the quantum information scrambling realization of the teleportation circuit.
Here, the encoding unitary $U(\alpha)=U_{\rm scr}(\alpha)$ is a scrambling matrix consisting of single and two-qubit matrices [cf. Figs.~\ref{fig:setup}(a) and (b) and Appendix~\ref{sect:appendix_numerical_details}].
%The scrambling distributes the information about the initial state across qubits (1-3). The shared entanglement between the encoding and decoding parties through the Bell pairs across qubits (2,5) and (3,4) makes this information available to the decoding party. The latter then decodes the scrambled information using the complex conjugate $U_{\rm scr}^*(\alpha)$. During the decoding, the initial state information reaches the target qubit 7 by the entanglement between qubits 6 and 7. Finally, the Bell measurement of the central qubits 3 and 4 projects the target qubit onto the desired input state.
The complexity and the non-locality of the scrambling circuit requires efficient scrambling and stable qubit coherence for successful teleportation.
Consequently, the teleportation fidelity~\eqref{eq:fidelity} increases with increasing scrambling strength $\alpha$ and decreases with increasing dephasing $\gamma$, see Fig.~\ref{fig:scrambling}(a).
We identify a regime of successful teleportation at strong scrambling and weak dephasing, outside of which the fidelity quickly decreases to $\overline{F}=1/2$ corresponding to a completely random target qubit state, $\rho_7=\one/2$.
The fidelity displays monotonous increase as a function of the scrambling strength at every dephasing strength, see Fig.~\ref{fig:scrambling}(b).
Conversely, the dephasing tampers with the teleportation success and decreases the fidelity.
%Evidently, scrambling and dephasing play a crucial role during the teleportation; only in the absence of noise and with highly efficient scrambling, the target qubit is correctly projected onto the source qubit's initial state, achieving a fidelity close to 1.
The fidelity's sensitivity to the dephasing noise results in an exponential decrease of the former as a function  of the dephasing strength $\gamma$, as detailed in Appendix~\ref{sect:appendix_fid(gamma)}.

In parallel to the fidelity, we expect the dephasing to affect the purity $\overline{\mathcal{P}}$ [cf. Eq.~\eqref{eq:purity}].
The dephasing leads to a loss of qubit coherence and a as such to an increasing classical mixture of the quantum states contributing to the final state.
Indeed, this effect leads to a monotonous decrease of purity as a function of the dephasing strength $\gamma$, see Fig.~\ref{fig:scrambling}(c).
More intriguing is the behavior of the purity as a function of the scrambling strength $\alpha$.
In the absence of dephasing, $\gamma=0$, the circuit is unitary and the state remains pure independent of $\alpha$.
For strong dephasing and efficient scrambling ($\alpha \lessapprox 1$), the purity approaches a minimal value of $\overline{\mathcal{P}}_{\rm min} = {1}/{2^5}$, see Fig.~\ref{fig:scrambling}(d).
This value can be explained by a completely dephased state $\rho=\1/2^7$ prior to the Bell measurement.
Then, the heralded Bell measurement projects the central qubits 3 and 4 onto pure states, while the rest of the qubits remain completely mixed.
Thus, the final state has the observed purity of $1/2^5$.
%On the other hand, at very strong dephasing, the final state's purity approaches the minimal value of $\overline{\mathcal{P}}_{\rm min} = {1}/{2^5}$, independent of $\alpha \to 1$.
%This purity corresponds to a completely dephased state of 5 qubits, $\rho=\one/2^5$, which shows that the circuit state before the Bell measurement must be close to $\one/2^7$, i.e., it is completely classical, too.
%Then, the projective Bell measurement projects qubits 3 and 4 onto a pure state, and the purity increases to the above value.

At intermediate dephasing strengths, the purity decreases with increasing scrambling strength.
This can be understood in the following way: the scrambling encodes and decodes the local information about the input qubit state across the whole circuit, leading to an increasing number of quantum states contributing to the final state superposition (in the computational basis).
Maintaining such a superposition requires a high degree of quantum coherence, making it increasingly vulnerable to dephasing.
This vulnerability leads to an exponential decay of the purity as a function of the dephasing strength $\gamma$, see Appendix~\ref{sect:appendix_fid(gamma)}.
So far, the noisy scrambling circuit behaves as expected: we find that both the fidelity and the purity show monotonous behavior as a function of the scrambling $\alpha$ as well as the dephasing $\gamma$. Crucially, we observe that scrambling tends to spread classical fluctuations, as it relies on a very specific fragile quantum states superposition.

In contrast to the fidelity and the purity, the system's excess entanglement displays an inversion of its behavior with the scrambling strength.
We quantify it using the logarithmic negativity~\eqref{eq: log negativity} between qubits 3 and 4 (i.e., subsystem $B$ consists of qubits 4-7) and identify two qualitatively different regimes as a function of the dephasing strength, see Fig.~\ref{fig:scrambling}(e).
Let us first discuss the excess entanglement of the circuit in absence of dephasing, $\gamma=0$.
The logarithmic negativity increases from $\overline{E}_\mathcal{N}=1$ at zero scrambling ($\alpha=0$) to $\overline{E}_\mathcal{N}=2$ at full scrambling ($\alpha=1$), corresponding to two Bell pairs across the bond, cf. Fig.~\ref{fig:scrambling}(f).
$\overline{E}_\mathcal{N}=1$ corresponds to the entanglement provided by the Bell pair between qubits 2 and 5 after the Bell measurement projects away the Bell pair between qubits 3 and 4, cf. Fig.~\ref{fig:setup}(a).
The increase of entanglement equivalent to a Bell pair for $\alpha=1$ is provided by the entangling gates contributing to $U_{\rm scr}$ and $U_{\rm scr}^*$ and is in line with the fact that the scrambling has to distribute the information non-locally.
This regime (I) exhibits a monotonous increase of the excess entanglement with increasing scrambling, and persists until a critical dephasing strength $\gamma_c \approx 0.038$, where the excess entanglement is almost independent of the scrambling parameter $\alpha$ [cf. Figs.~\ref{fig:scrambling}(e) and (f)].
Conversely, at higher dephasing strengths, $\gamma \geq \gamma_c$, we find a regime (II), where the excess entanglement decreases as a function of the scrambling.
As we will see below, in this regime, the noise forces the Bell measurement to drain an increasing amount of the scrambled entanglement for teleportation.
As such, this effect can be understood as a scrambling-induced entanglement suppression.
The identification of two qualitatively different regimes in the behavior of entanglement in a scrambling circuit is the main result of our work.
Noticeably, a least mean square fit to the experimental data of Ref.~\cite{landsman_verified_2019} reveals a realistic dephasing stength of $0.01\lessapprox \gamma \lessapprox 0.02$, i.e., current quantum devices operate close to the noisy regime of scrambling-induced entanglement suppression.
In the following, we will compare our findings to the SWAP-gate-based teleportation circuit to confirm that the entanglement suppression can indeed be attributed to quantum information scrambling.
\section{Comparison with SWAP-gate-based quantum teleportation}
\label{sec: comparison}
In order to single out the effects of quantum information scrambling, we compare the above findings to a SWAP-gate-based teleportation circuit, which for comparison reasons is extended from its minimal 3 qubit size to 7 qubits, see Figs.~\ref{fig:setup}(a) and (c).
Instead of scrambling the initial state information across the whole setup, this circuit enables teleportation by local exchange of information using SWAP gates~\cite{yoshida2019disentangling}.
Then, $U(\alpha)=U_{\rm SWAP}$ consists of two consecutive parametrized SWAP gates, where $0\leq \alpha\leq 1$ is the swapping strength ranging from no swap at all ($\alpha=0$) to perfect qubit exchange ($\alpha=1$) [cf. Appendix~\ref{sect:appendix_numerical_details}].
We recall that the minimal teleportation protocol requires a single shared Bell pair between the encoding and decoding parties, such that for $\alpha=1$ our setup has two excess Bell pairs untouched by the Bell measurement at the end of the protocol.
This is a first main difference to the teleportation by scrambling, where all 3 Bell pairs are necessary to scramble the information across the circuit.
As for the scrambling circuit, we consider the teleportation fidelity~\eqref{eq:fidelity}, the purity~\eqref{eq:purity}, and the logarithmic negativity~\eqref{eq: log negativity} as functions of the parameter $\alpha$ and the dephasing strength $\gamma$.

\begin{figure}[t]
    \centering
    \includegraphics{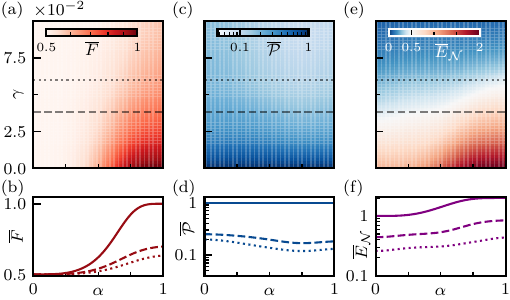}
    \caption{Fidelity $\overline{F}$, purity $\overline{\mathcal{P}}$, and negativity $\overline{E}_\mathcal{N}$ of the final state of the SWAP-gate-based teleportation circuit [cf. Figs.~\ref{fig:setup}(a) and (c)] as a function of scrambling strength $\alpha$ and dephasing strength $\gamma$.
    (b,d,f) show the same data as (a,c,e) for specific dephasing strength cuts $\gamma \in \{0, 0.038, 0.06\}$ (solid, dashed, dotted).
    (a) High teleportation fidelity ($\overline{F} \approx 1$) is achieved with high swapping strength $\alpha$ and minimal dephasing. Weak $\alpha$ and/or strong dephasing lead to a fidelity $\overline{F} \approx 1/2$, indicating an unknown final state. 
    (b) The fidelity increases monotonically with the swapping strength $\alpha$ for all dephasing strengths. 
    (c) Increasing dephasing noise $\gamma$ results in a more classical final state characterized by low purity $\overline{\mathcal{P}}$. 
    (d) In the absence of dephasing, the final state is pure independent of $\alpha$.
    For $\gamma>0$, the purity in general decreases with increasing swapping strength $\alpha$, up to a slight increase at high values of $\alpha$. 
    (e) The logarithmic negativity $\overline{E}_\mathcal{N}$ is suppressed by dephasing and enhanced by increasing swapping strength $\alpha$.
    (f) Across all dephasing strengths $\gamma$, the logarithmic negativity increases monotonously with the scrambling parameter $\alpha$.}
    \label{fig:swap}
\end{figure}

Similar to the scrambling circuit, weak dephasing and efficient swaps are indispensable for successful teleportation, see Fig.~\ref{fig:swap}(a).
Still, we find another difference to the scrambling circuit: whereas scrambling delocalizes the information across many qubits, the SWAP-gate-based teleportation distributes the quantum information more locally.
Consequently, the SWAP circuit relies on more local quantum coherence, making it in turn less sensitive to the dephasing noise.
This is evident in a more extended regime of high fidelities towards stronger dephasing [cf. Fig.~\ref{fig:swap}(a) compared with Fig.~\ref{fig:scrambling}(a)].
Apart from the decreased sensitivity, the fidelity displays the same monotonous behavior as in the scrambling circuit.
In the absence of dephasing, it increases from $\overline{F}=1/2$ for $\alpha=0$ corresponding to a random final state $\rho_7=\one/2$ to $\overline{F}=1$ for $\alpha=1$ signalling perfect teleportation, see Fig.~\ref{fig:swap}(b).
For $\gamma>0$, the increase remains monotonous but gets damped.

Next, we discuss the purity~\eqref{eq:purity} of the final state to quantify how classical it is.
In conjunction with the fidelity behavior, the purity decreases with increasing noise but shows lower sensitivity to the latter compared with the scrambling circuit [cf. Fig.~\ref{fig:swap}(c)].
As expected, for $\gamma=0$ the final state remains pure ($\overline{\mathcal{P}}=1$) independent of the swapping strength $\alpha$.
The lower sensitivity to noise compared with the scrambling circuit leads to a decline in purity occurring at a significantly slower rate, see Fig.~\ref{fig:swap}(d).
Crucially, in contrast to the scrambling circuit, we find a non-monotonous behavior of the purity with the swapping strength $\alpha$.
After an initial decrease, the purity assumes a minimal value followed by a slight increase when $\alpha$ approaches 1.
This is a consequence of our SWAP gate parametrization: at intermediate values of $\alpha$, the SWAP gate generates a coherent superposition of the qubit states (in the computational basis).
This superposition is highly susceptible to the dephasing, whereas at very low or very high $\alpha$, the swap is either absent or fast and efficient.
Then, the qubits spend less time in a superposition and are more stable against dephasing.

Finally, we consider the excess entanglement for a bipartition of the circuit between qubits 3 and 4.
Intriguingly, we do not find two regimes as we observed in the scrambling circuit: increasing the swapping strength $\alpha$ always leads to more excess entanglement, see Fig.~\ref{fig:swap}(e).
Only for weak dephasing, the two circuits show a very similar behavior.
In the absence of noise, also the SWAP circuit's excess entanglement increases from $\overline{E}_\mathcal{N} = 1$ for $\alpha=0$ to $\overline{E}_\mathcal{N} = 2$ for $\alpha=1$.
In contrary to the scrambling circuit, however, this monotonous behavior pertains at all dephasing strengths, albeit in a damped way.
We conclude that the inversion of the excess entanglement scaling is a consequence of quantum information scrambling.
However, it is not yet clear which part of the circuit leads to such scrambling-suppressed entanglement. 
To answer this question, we consider in the following the total entanglement change due to the information distribution by $U(\alpha)$ (and $U^*(\alpha)$)  and the Bell measurement separately.
\section{Total entanglement}
\label{sec: total entanglement}
In this Section, we compare the entanglement behavior between the scrambling and the SWAP-gate-based teleportation circuit in more detail.
To this end, we consider the change in total entanglement in the circuit during the application of the unitaries $U=U_{\rm scr},\,U_{\rm SWAP}$ and their conjugates [cf. Fig.~\ref{fig:setup}(a)],
\begin{equation}
\label{eq: delta E U}
    \Delta\overline{E}_{\mathcal{N},U} = \overline{E}_{\mathcal{N},{\rm tot}}(t_2) - \overline{E}_{\mathcal{N},{\rm tot}}(t_1)\,,
\end{equation}
where $\overline{E}_{\mathcal{N},{\rm tot}}(t)$ is the total entanglement (the sum over the logarithmic negativity over all bonds) at time $t$, and $t_1$ and $t_2$ are the times right before and after the application of $U$ and $U^*$.
Furthermore, we consider the total entanglement change during the Bell measurement,
\begin{equation}
\label{eq: delta E M}
    \Delta\overline{E}_{\mathcal{N},M} = \overline{E}_{\mathcal{N},{\rm tot}}(t_3) - \overline{E}_{\mathcal{N},{\rm tot}}(t_2)\,,
\end{equation}
where $t_3$ is the time right after the Bell measurement.
The total entanglement is better suited to observe the overall effect of scrambling and swapping than the entanglement between qubits 3 and 4 because the latter is only indirectly affected by the unitaries $U$.
Note that the total entanglement cannot distinguish between the generation and the delocalization of entanglement.

\begin{figure}[t]
    \centering
    \includegraphics{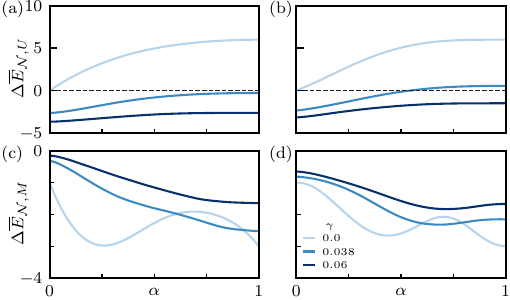}
    \caption{The change in the total logarithmic negativity $\Delta\overline{E}_{\mathcal{N},U}$ [cf. Eq.~\eqref{eq: delta E U}] during the (a) scrambling  and (b) swapping process and the change in the total logarithmic negativity $\Delta\overline{E}_{\mathcal{N},M}$ [cf. Eq.~\eqref{eq: delta E M}] by the projective Bell measurement for the (c) scrambling and (d) SWAP-gate-based circuit as a function of the transmission strength $\alpha$ for different dephasing strengths $\gamma$.
    (a) Scrambling generates entanglement while dephasing destroys it, leading to a net negative entanglement production at high dephasing strengths.
    (b) Swapping distributes entanglement while dephasing destroys it, leading to a net negative entanglement production at high dephasing strengths. The swapping is slightly less susceptible to the noise as the scrambling in (a). 
    (c) Stronger scrambling leads to more entanglement consumption of the Bell measurement.
    At weak dephasing, the delocalization of the initial Bell pair entanglement leads to a non-monotonous behavior of $\Delta\overline{E}_{\mathcal{N},M}$.
    (d)  Stronger swapping in general leads to more entanglement consumption (smaller $\Delta\overline{E}_{\mathcal{N},M}$) of the Bell measurement.
    At strong swapping strengths, there is again a slight increase in $\Delta\overline{E}_{\mathcal{N},M}$.
    As in (c), at weak dephasing, the delocalization of the initial Bell pair entanglement leads to a non-monotonous behavior of $\Delta\overline{E}_{\mathcal{N},M}$.}
    \label{fig:entanglement_diff}
\end{figure}

First, we consider the total entanglement change~\eqref{eq: delta E U} due to the information distribution for both the scrambling and the SWAP-gate-based teleportation circuit, see Figs.~\ref{fig:entanglement_diff}(a) and (b).
In the absence of dephasing, $\gamma=0$, $\Delta\overline{E}_{\mathcal{N},U}$ increases from $\Delta\overline{E}_{\mathcal{N},U}=0$ for $\alpha=0$ (because then $U(0)=\one$) to a finite value of $\Delta\overline{E}_{\mathcal{N},U}=6$ at $\alpha=1$ for both realizations.
When $\gamma>0$, there is a competition between entanglement enhancement by $U$ and $U^*$ and entanglement destruction due to dephasing.
Interestingly, the dephasing strength that exactly destroys the total entanglement generated by the scrambling for $\alpha=1$ [cf. Fig.~\ref{fig:entanglement_diff}(a)] coincides with the critical dephasing strength $\gamma_c$ separating the two entanglement-scaling regimes of the scrambling circuit [cf. Sec.~\ref{sec: scrambling} and Fig.~\ref{fig:scrambling}(e)].
The SWAP-gate-based teleportation circuit is slightly more stable against dephasing.
At high dephasing strengths, the noise overpowers the information distribution and the net total entanglement change during the information distribution process is negative at all scrambling strengths $\alpha$ for both circuit realizations.

We recall that for the scrambling circuit, the regime (II) at high dephasing strengths was identified by an inverse behavior of the excess entanglement with the scrambling strength $\alpha$, i.e., it decreased with increasing scrambling strength [cf. Fig.~\ref{fig:scrambling}(e)].
Given the monotonous increase of the generated total entanglement $\Delta\overline{E}_{\mathcal{N},U}=0$ during the scrambling, this behavior can only be explained in combination with the entanglement~\eqref{eq: delta E M} destroyed during the Bell measurement at the end of the circuit.
Indeed, we find that in general, the entanglement consumption of the Bell measurement increases for both the scrambling and the SWAP-gate-based teleportation circuit with increasing $\alpha$, see Figs.~\ref{fig:entanglement_diff}(c) and (d).

Interestingly, both circuits show a non-monotonous behavior of $\Delta\overline{E}_{\mathcal{N},M}$ at very weak noise.
We attribute this to competing local and global effects of the information distribution and the consecutive Bell measurement: for $\alpha=1$, the information distribution has no effect, and the Bell measurement simply projects out the Bell pair initialized between qubits 3 and 4, hence $\Delta\overline{E}_{\mathcal{N},M}=-1$.
By increasing $\alpha$, the qubits 3 and 4 become correlated with the rest of the system, and the Bell measurement can consume more entanglement than just the single initial central Bell pair.
Upon further increase of $\alpha$, the Bell pair entanglement of the qubit pairs (2,5) and (3,4) gets more delocalized and less susceptible to the local Bell measurement, leading to an increase of $\Delta\overline{E}_{\mathcal{N},M}$.
Finally, at high values of $\alpha$, also the Bell pair entanglement between qubits 6 and 7 is distributed across the system and therefore affected by the Bell measurement, too, which again leads to a decrease of $\Delta\overline{E}_{\mathcal{N},M}$.

At higher dephasing strengths, this behavior is suppressed, and the Bell measurement in general consumes more entanglement with stronger information distribution, see Figs.~\ref{fig:entanglement_diff}(c) and (d).
Note that for the SWAP-gate-based teleportation circuit, $\Delta\overline{E}_{\mathcal{N},M}$ increases again at high values of $\alpha$.
This is in line with the observed increase in purity, which can be explained by the entangling nature of the SWAP gate at intermediate swapping strengths [cf. Sec.~\ref{sec: comparison} and Figs.~\ref{fig:swap}(c) and (d)].
This difference in $\Delta\overline{E}_{\mathcal{N},M}$ between the scrambling and the SWAP-gate-based circuit at high transmission strengths $\alpha$ together with the higher noise-susceptibility of the scrambling circuit explains why we observe two regimes of entanglement scaling in the scrambling circuit but not in the traditional teleportation circuit: in the scrambling circuit, the generated entanglement $\Delta\overline{E}_{\mathcal{N},U}$ is highly suppressed at high scrambling strengths $\alpha$, and at the same time, the entanglement consumption $\Delta\overline{E}_{\mathcal{N},M}$ is increased.
This confirms the scrambling-induced entanglement suppression in noisy quantum circuits.

\section{Concluding remarks}
\label{sec: conclusion}
We have analyzed the effect of noise in an experimentally realized teleportation circuit serving as a testbed for quantum information scrambling in black holes~\cite{landsman_verified_2019}.
Thereby, we have found a surprising effect of scrambling in the presence of noise: through the delocalization of information, it makes a system more and more vulnerable to dephasing.
As a consequence, a scrambling-based teleportation protocol needs to consume more entanglement during the Bell measurement step to faithfully teleport information.
This process can be understood as a scrambling-induced entanglement suppression in the presence of noise.
We have compared the scrambling circuit to a SWAP-gate-based teleportation circuit, where the information about the initial state is transported in a more local fashion than in the scrambling circuit.
For SWAP-gate-based teleportation, we do not observe entanglement suppression.

Our findings shed light into the complex process of relaxation and thermalization in quantum systems: the interplay between quantum information scrambling and entanglement during thermalization is highly affected by the presence of noise.
At weak noise, they mutually increase while with strong noise, the scrambling suppresses the entanglement.
Thermalization in quantum systems has been experimentally analyzed in a variety of setups~\cite{Kinoshita_Wenger_Weiss_2006,doi:10.1126/science.1224953,Trotzky_Chen_Flesch_McCulloch_Schollwoeck_Eisert_Bloch_2012,PhysRevLett.117.170401,PhysRevLett.117.170401,Neill_Roushan_Fang_Chen_Kolodrubetz_Chen_Megrant_Barends_Campbell_Chiaro_et_al._2016,doi:10.1126/science.abl6277,PRXQuantum.4.020318,doi:10.1126/science.aaf6725}.
Moreover, recent advances in the experimental observation of entanglement~\cite{Friis_Vitagliano_Malik_Huber_2019,Kokail_vanBijnen_Elben_Vermersch_Zoller_2021,Joshi_Kokail_van_Bijnen_Kranzl_Zache_Blatt_Roos_Zoller_2023,Islam_Ma_Preiss_Eric_Tai_Lukin_Rispoli_Greiner_2015} offer exciting perspectives to gain new insights into the scrambling and entanglement dynamics of quantum systems.
Furthermore, the fact that we observe the scrambling-induced entanglement suppression on a 7-qubit quantum circuit has consequences for computation on today's noisy intermediate-scale quantum (NISQ) devices~\cite{preskill_2018,brooks_2019}: depending on the strength of the noise, it is either favorable to scramble information across the circuit or distribute it in a more local manner, e.g. using SWAP gates.
This also implies that measurement-based quantum computation~\cite{jozsa2005introduction,briegel2009measurement} is only possible for extremely noise-resistent devices.
Finally, our findings suggest that a black hole's ability to generate entanglement by scrambling is severely limited by the decohering effect of its Hawking radiation.

\begin{acknowledgments}
The authors acknowledge financial support by ETH Research Grant ETH-51 201-1 and the Deutsche Forschungsgemeinschaft (DFG) - project number 449653034 and through SFB1432, as well as from the Swiss National Science Foundation (SNSF) through the Sinergia Grant No.~CRSII5\_206008/1.
\end{acknowledgments}

\appendix
\section{Explicit form of the teleportation circuits \label{sect:appendix_circuits}}

\begin{figure*}[ht]
    \centering
    \includegraphics{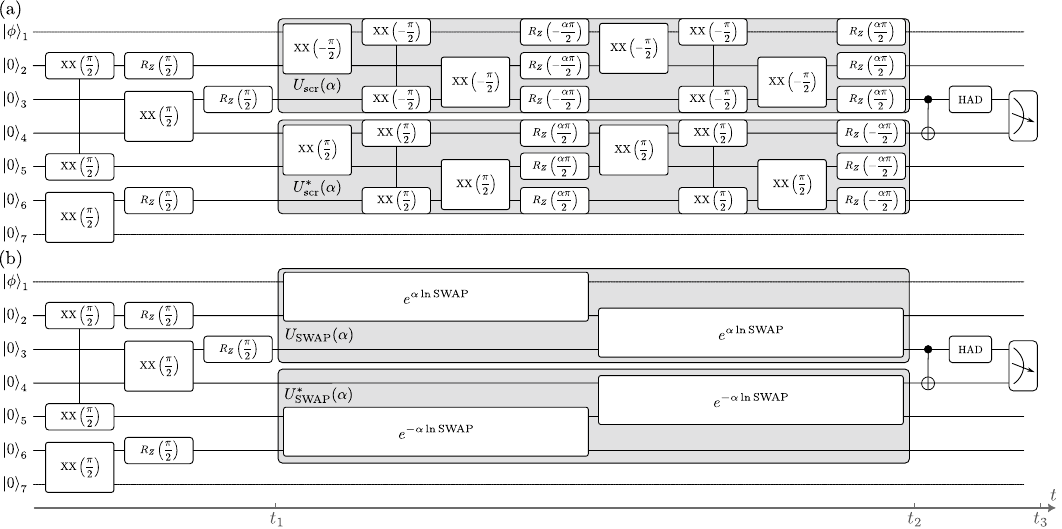}
    \caption{The 7-qubit (a) scrambling and (b) SWAP-gate-based teleportation circuit [c.f. Figs.~\ref{fig:setup}(a), (b) and (c)].
    For both circuits, Bell pairs between qubits (2,5), (3,4) and (6,7) are created during $t \in [0, t_1]$ by application of ${\rm XX}(\pi/2)$ and $R_Z(\pi/2)$ [cf. Eqs.~\eqref{eq: XX} and~\eqref{eq: RZ}].
    (a) In the scrambling circuit of Ref.~\cite{landsman_verified_2019}, the scrambling unitaries $U_{\rm scr}$ and $U^*_{\rm scr}$ are applied during $t \in [t_1, t_2]$.
    The scrambling strength $\alpha \in [0, 1]$ controls the $Z$-rotations.  
    (b) In the SWAP-gate-based circuit, $U_{\text{SWAP}}(\alpha)$ and $U^*_{\text{SWAP}}(\alpha)$ consist of parameterized SWAP gates $\exp[\pm\alpha \ln SWAP]$ [cf. Eq.~\eqref{eq:swap}], where $\alpha$ is the swapping strength.
    Lastly, in both circuits and for $t \in [t_2, t_3]$, a projective Bell measurement is performed on qubits 3 and 4 by application of CNOT and Hadamard gates [cf. Eqs.~\eqref{eq: CNOT} and~\eqref{eq: HAD}] and a projective measurement of both qubits in the $Z$-basis.}
    \label{fig:circuits_gates}
\end{figure*}

Here, we provide details on the realizations of the scrambling and the SWAP-gate-based teleportation circuit.
The scrambling circuit of Ref.~\cite{landsman_verified_2019} is shown in Fig.~\ref{fig:circuits_gates}(a) and consists of the following single-qubit and two-qubit quantum gates~\cite{nielsen_chuang_2010},
\begin{align}
\label{eq: XX}
    {\rm XX}(\phi) &= \exp\left[\frac{i\phi}{2} X \otimes X\right]\,, \\
    \label{eq: RZ}
    R_Z(\phi)&=\exp\left[\frac{i\phi}{2} Z\right]\,, \\
    \label{eq: CNOT}
    {\rm CNOT} &= \exp\left[\frac{i\pi}{4} (\one - Z) \otimes (\one - X))\right]\,, \\
    \label{eq: HAD}
    {\rm HAD} &= \exp\left[\frac{i \pi}{2\sqrt{2}} (X + Z)\right]\,,
\end{align}
where $X$ and $Z$ are the Pauli-X and Pauli-Z matrices, respectively, and ${\rm HAD}$ is the Hadamard gate.
Bell pairs are created by consecutive application of the ${\rm XX}(\pi/2)$ interaction and the $R_Z(\pi/2)$ rotation gate.
The scrambling unitary $U_{\rm scr}(\alpha)$ is implemented using a sequence of ${\rm XX}(-\pi/2)$ and $R_Z(\pm\alpha\pi/2)$ gates, with the scrambling strength $\alpha$.
For $\alpha=0$, the unitary reduces to $U_{\rm scr}(0)=\1$.
Finally, the Bell measurement involves a ${\rm CNOT}$ and a Hadamard gate.
Note that the conjugate scrambling unitary $U^*_{\rm scr}(\alpha)$ is achieved by negative rotation for all ${\rm XX}$ and $R_Z$ gates compared to $U_{\rm scr}(\alpha)$.

The SWAP-gate-based teleportation circuit has equivalent Bell pair creation and Bell measurement steps involving the gates~\eqref{eq: XX}-\eqref{eq: HAD} as the scrambling circuit, see Fig.~\ref{fig:circuits_gates}(b).
Instead of the scrambling unitaries $U_{\rm scr}$ and $U^*_{\rm scr}$, however, the SWAP-gate-based circuit realizes teleportation using parametrized SWAP gates $\exp[\pm\alpha \ln {\rm SWAP}]$, where the logarithm of the SWAP gate is given by
\begin{equation}
    \ln {\rm SWAP} = \frac{i\pi}{2} \begin{pmatrix}
0 & 0 & 0 & 0\\
0 & 1 & -1 & 0\\
0 & -1 & 1 & 0\\
0 & 0 & 0 & 0 \label{eq:swap}
\end{pmatrix},
\end{equation}
such that for $\alpha=1$, it holds $\exp[ \ln {\rm SWAP}]={\rm SWAP}$.

We use continuous implementations of the gates, i.e., we apply them over a time interval $t\in[\tau_0, \tau_0+\tau]$, where $\tau_0$ is the start time of the gate in the circuit, and we choose a gate application time of $\tau=1$.
For example, to achieve the $R_Z(\pi/2)$-rotation, we apply the time-dependent gate
\begin{equation}
\label{eq: continuous gate}
    G(t) = \exp\left[\frac{i\pi t}{4\tau} Z\right]\,, \quad t\in [0, \tau]\,,
\end{equation}
where we have set $\tau_0=0$ for simplicity.
The continuous-in-time implementations of the other gates follow analogously.
Each gate is applied over a time $\tau=1$ except for the parametrized SWAP gates, which we apply over a time $\tau'=4$ to achieve equivalent duration for both teleportation circuits.
The exponential form of the continuous gate~\eqref{eq: continuous gate} permits to think of the gate application as the time evolution governed by a Hamiltonian $H$,
\begin{equation}
    G(t) = \exp\left[-\frac{i}{\hbar} Ht\right]\,,\quad H=-\frac{\hbar \pi}{4\tau} Z\,.
\end{equation}
The Hamiltonian $H(\alpha,t)$ appearing in the Lindblad master equation~\eqref{eq:lindblad} describes such continuous gate implementations, and the time-dependence reflects the fact that the applied gates change over time during the circuit.

\section{Details on the numerical implementation}\label{sect:appendix_numerical_details}

In the main body, we have introduced the Lindblad master equation~\eqref{eq:lindblad}.
Here, we explain how we numerically evolve the Lindblad equation in time.
We make use of the superoperator formalism~\cite{breuer_petruccione_2002} to integrate Eq.~\eqref{eq:lindblad} and arrive at
\begin{equation}
    \rho(t) = \mathcal{T}\exp\left[\int_0^t\mathcal{L}\left(t^\prime\right)dt^\prime\right][\rho(0)]\,,
    \label{eq:integrated_lindblad_master}
\end{equation}
where $\mathcal{T}$ is the time ordering operator, $\rho(0)$ the initial state of the circuit, and $\mathcal{L}(t)$ the Liouvillian superoperator~\cite{breuer_petruccione_2002}
\begin{equation}
    \mathcal{L}(t) = \mathcal{L}_H(t) + \mathcal{D}\,.
    \label{eq: liou}
\end{equation}
The first term of the Liouvillian~\eqref{eq: liou} corresponds to the unitary evolution,
\begin{equation}
    \mathcal{L}_H(t) = -\frac{i}{\hbar}\left(H(\alpha, t) \otimes \one - \mathbb{1} \otimes H(\alpha, t)^*\right)\,,
\end{equation}
where $H(\alpha, t)$ is the Hamiltonian describing the quantum gates in the circuits and $\otimes$ denotes the Kronecker product.
The second term of the Liouvillian~\eqref{eq: liou} is a sum over qubit-local dissipators $\mathcal{D}_i$,
\begin{equation}
    \begin{split}
        \mathcal{D} &= \sum_{i=1}^7 \mathcal{D}_i \\
        &=\sum_{i=1}^7 \gamma \left[ n_i \otimes n_i - \frac{1}{2}\left( \one \otimes n_i+ n_i \otimes \one \right)\right]\,,
    \end{split}
\end{equation}
where $\gamma$ is the dephasing strength and $n_i = \frac{1}{2} (\one + Z_{i})$, with $Z_{i}$ the Pauli-Z matrix on site $i$.

Next, we discretize the time evolution into time bins $\Delta t$ and perform a Trotter-Suzuki decomposition~\cite{trotter_product_1959,suzuki_generalized_1976} of the exponentiated Liouvillian,
\begin{equation}
    \exp\left[\Delta t \left(\mathcal{L}_H(t) + \mathcal{D}\right)\right] \approx \exp[\Delta t \mathcal{L}_H(t)] \exp[\Delta t \mathcal{D}]\label{eq:trotter-suzuki}\,,
\end{equation}
i.e., we perform the unitary and dissipative updates separately and omit correction terms of $\mathcal{O}(\Delta t^2)$ and higher.
The unitary update is then achieved by
\begin{align}
\label{eq: unitary update}
    \exp[\Delta t \mathcal{L}_H(t)][\rho] = \tilde{U}(t)\rho \tilde{U}^\dag(t)\,,
\end{align}
where $\tilde{U}(t)=\exp[-iH(\alpha,t)\Delta t/\hbar]$ is the unitary time evolution operator.
We use exact diagonalization for the update~\eqref{eq: unitary update}, i.e., we numerically represent the unitary $\tilde{U}(t)$ without further approximations.

For the dissipative update, we use a Kraus operator sum representation~\cite{kraus_general_1971,nielsen_chuang_2010},
\begin{equation}
\label{eq: kraus op sum}
    \exp\left[\Delta t \sum_{i=1}^7 \mathcal{D}_i\right] [\rho] = \prod_i \sum_{j=1}^2 {K_j}_i\ \rho\ {K_j}_i^\dag\,,
\end{equation}
with local Kraus operators
\begin{equation}
\begin{split}
     & {K_1}_i = \one + n_i\ \left(e^{- \gamma \Delta t} - 1\right) \\
    & {K_2}_i = \sqrt{1 - e^{-2 \gamma \Delta t}}\ n_i.
\end{split}
\end{equation}
Note that the separation in~\eqref{eq: kraus op sum} is exact as the local Kraus operators commute.

As stated in the main text, we average every circuit run over several initial states.
Specifically, we repeat the circuit for initial states $\ket{\phi} \in \{\ket{0}_X, \ket{1}_X, \ket{0}_Y, \ket{1}_Y, \ket{0}_Z, \ket{1}_Z,\}$, where $\ket{0(1)}_i$ is the positive (negative) eigenstate of the Pauli-$i$ matrix.

\section{The effect of dephasing}\label{sect:appendix_fid(gamma)}

\begin{figure}[hb]
    \centering
    \includegraphics{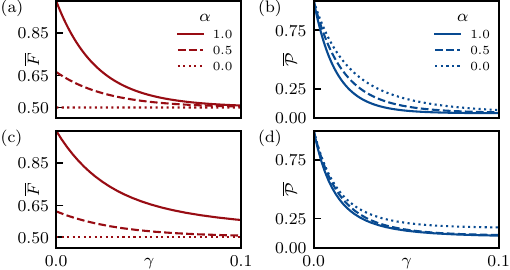}
\caption{The fidelity $\overline{F}$ [cf. Eq.~\eqref{eq:fidelity}] and purity $\overline{\mathcal{P}}$ [cf. Eq.~\eqref{eq:purity}] as a function of the error strength $\gamma$ for $\alpha \in \{0, 0.5, 1\}$ (dotted, dashed, solid). (a)-(b) In the scrambling circuit [c.f. Fig.~\ref{fig:setup}(b) and Fig.~\ref{fig:circuits_gates}(a)], both the fidelity and the purity exponentially decay with increasing $\gamma$.
For $\alpha=0$, the fidelity is not affected by dephasing as the target qubit state $\rho_7$ remains random in the absence of any scrambling.
(c)-(d) The SWAP-gate-based teleportation circuit  [c.f. Fig.~\ref{fig:setup}(c) and Fig.~\ref{fig:circuits_gates}(b)] shows a similar exponential decay of the fidelity and the purity as the scrambling circuit in (a) and (b). However, the fidelity is less sensitive to noise for $\alpha=1$ and the purity decays to a slightly higher value for large values of $\gamma$.
}
    \label{fig:against_gamma}
\end{figure}

Here, we investigate the effect of dephasing on the average teleportation fidelity $\overline{F}$ [cf. Eq.~\eqref{eq:fidelity}] and the average circuit end state purity $\overline{\mathcal{P}}$ [cf. Eq.~\eqref{eq:purity}].
For both circuit realizations, the fidelity and the purity in general decay exponentially as a function of the dephasing strength $\gamma$, see Fig.~\ref{fig:against_gamma}.
For no information transmission, i.e., $\alpha=0$, both the scrambling circuit's and the SWAP-gate-based circuit's target qubit end state remains in a random state, $\rho_7=\one/2$, and the fidelity is independent of the dephasing [cf. Figs.~\ref{fig:against_gamma}(a) and~\ref{fig:against_gamma}(c)].
In line with the discussions in the main text, the fidelity in the scrambling circuit is more sensitive to the dephasing than the fidelity in the SWAP-gate-based circuit [cf. Fig.~\ref{fig:against_gamma}(a) vs. Fig.~\ref{fig:against_gamma}(c) for $\alpha=1$].
At strong dephasing noise, the average purity of the scrambling circuit decays to a lower value compared to the SWAP-gate-based circuit, compare Fig.~\ref{fig:against_gamma}(b) to Fig.~\ref{fig:against_gamma}(d).
This can be explained by the smaller number of gates in the SWAP-gate-based circuit: e.g., for $\alpha=0$, there is no gate application between $t_1$ and $t_2$ for the SWAP-gate-based circuit, and consequently less superposition and coherence destroyed by dephasing.

\bibliography{apssamp}% Produces the bibliography via BibTeX.

\end{document}